\documentclass[]{spie}  
\usepackage[]{graphicx}
\usepackage[usenames,dvipsnames]{color}
\usepackage{amsmath,amssymb}

\newcommand{\green}{}

\newcommand{\arcsec}{\mbox{$^{\prime\prime}$}}
\newcommand{\farcs}{\mbox{$\!\!^{\prime\prime}$}}

\newcommand{\f}{\emph{f}/}

\title{HexPak and GradPak: variable-pitch dual-head IFUs for the WIYN 3.5m Telescope Bench Spectrograph}

\author{Corey M.\ Wood\supit{a}, Matthew A.\ Bershady\supit{a}, Arthur D.\ Eigenbrot\supit{a}, Scott A.\ Buckley\supit{a}, John S.\ Gallagher III\supit{a}, Eric J.\ Hooper\supit{a}, Andrew I.\ Sheinis\supit{b}, Michael P.\ Smith\supit{a}, Marsha J.\ Wolf\supit{a}
\skiplinehalf
\supit{a}University of Wisconsin--Madison, 475 N.\ Charter St., Madison, WI, USA; \\
\supit{b}Australian Astronomical Observatory, P.O.\ Box 296, Epping, NSW 1710, Australia
}

\authorinfo{Further author information:\\
C.M.W.: E-mail: wood@astro.wisc.edu, Telephone: +1 (608) 262--3071}

\begin{document} 
  \maketitle 

\begin{abstract}
We describe the design, construction, and expected performance of two new fiber integral field units (IFUs) --- HexPak and GradPak --- for the WIYN 3.5m Telescope Nasmyth focus and Bench Spectrograph.
These are the first IFUs to provide formatted fiber integral field spectroscopy with simultaneous sampling of varying angular scales.
HexPak and GradPak are in a single cable with a dual-head design, permitting easy switching between the two different IFU heads on the telescope without changing the spectrograph feed: the two heads feed a variable-width double-slit.
Each IFU head is comprised of a fixed arrangement of fibers with a range of fiber diameters.
The layout and diameters of the fibers within each array are scientifically-driven for observations of galaxies: HexPak is designed to observe face-on spiral or spheroidal galaxies while GradPak is optimized for edge-on studies of galaxy disks.
HexPak is a hexagonal array of 2.9 arcsec fibers subtending a 40.9 arcsec diameter, with a high-resolution circular core of 0.94 arcsec fibers subtending 6 arcsec diameter.
GradPak is a 39 by 55 arcsec rectangular array with rows of fibers of increasing diameter from angular scales of 1.9 arcsec to 5.6 arcsec across the array.
The variable pitch of these IFU heads allows for adequate sampling of light profile gradients while maintaining the photon limit at different scales.
\end{abstract}



\section{INTRODUCTION}
\label{sec:intro}
HexPak and GradPak are two new formatted fiber optic integral field units (IFUs) for the WIYN 3.5m telescope Bench Spectrograph\footnotemark.
\footnotetext{The WIYN Observatory is a joint facility of the University of Wisconsin-Madison, Indiana University, Yale University, and the National Optical Astronomy Observatory.}
These two IFUs are unique because they are the first formatted fiber IFUs to include multiple fiber diameters within the same fiber head.
Including multiple fiber diameters allows each of these IFUs to simultaneously sample different angular scales within the same observation.
The smaller fibers can gather light from higher surface brightness regions (e.g.\ the core or midplane of a galaxy disk) while the larger fibers can collect light from fainter, more diffuse regions (e.g.\ larger scale heights or scale radii of a disk), thereby enabling high S/N measurements to be obtained simultaneously at a range of spatial positions.

The two IFUs share the same cable and ``foot'' for mounting onto the WIYN Bench Spectrograph.
Sharing the same cable minimizes the total volume required for routing within the WIYN telescope in an already over-filled system.
Sharing the same foot results in a unique dual-slit design, allowing the two heads to be exchanged at the telescope without requiring changes to the spectrograph system.
HexPak has a high-resolution core of fibers three times smaller in diameter than the surrounding fibers.
As a hexagonal array with a circular, high-resolution core, HexPak is tailored for studies of radially-distributed, diffuse light sources, such as face-on galaxy disks, spheroidal galaxies, or star clusters.
The GradPak head consists of five different fiber sizes, arranged in rows to form a gradient of fiber diameters from one edge of the array to the other.
It is designed for integral field spectroscopy of edge-on galaxies, making it well-suited for studying extra-planar gas and stars in spiral galaxy disks.

Including multiple fiber diameters comes at a cost, however.
In the case where the system spectral resolution is limited by the slit width, this will result in a varying spectral resolution, inversely proportional to the fiber diameter.
In this case the maximum resolution will change by a factor of 3 for GradPak and 3.1 for HexPak, increasing from the largest to the smallest fibers.
However, the smallest reimaged fiber sizes will have contributions from optical aberrations from the spectrograph.
As a result, we expect the achievable resolution to increase only by a factor of 2--2.5 for HexPak.

The science impact of changing spectral resolution depends on the specific application.
For example, the velocity dispersion of stars is expected to increase with scale height above the disk midplane in edge-on disk galaxies.
For the study of stellar velocity dispersions in spheroidal or face-on disk galaxies, as another example, it would be advantageous to \emph{increase} spectral resolution with radius, since systems become dynamically colder moving outward.
This is opposite what these instruments deliver.
On the other hand, at lower surface brightness the limits of S/N prevent useful information from being obtained at high spectral resolution, and in this sense these instruments provide a practical balance between signal and resolution.
As we describe below, ample sky fibers are included for all fiber sizes to ensure excellent sky subtraction.

This instrument follows in the legacy of the excellent WIYN fiber IFUs DensePak\cite{Barden98} and SparsePak\cite{Bershady04,Bershady05}.
The primary impetus behind this project was the increased throughput and image quality of the newly redesigned WIYN Bench Spectrograph\cite{Barden94,Bershady08,Knezek10}.
In the process an opportunity arose to rebuild the decommissioned DensePak IFU.
The fiber from DensePak is being reused for the larger fibers in the new HexPak array, and most of the hardware in the cable ``foot'' housing that terminates the cable in the spectrograph room at WIYN is reused from the DensePak cable.
HexPak contains additional fibers that were newly purchased for this project.
GradPak is made entirely using new fiber.
Additionally, all the cabling and head mount hardware is newly constructed.

In \S\ref{sec:design} we detail the key science drivers that served as a design target for the instrument, as well as describe the design challenges inherent in the design of formatted IFUs with multiple fiber diameters.  We detail the construction process in \S\ref{sec:construction} and summarize the project in \S\ref{sec:conclusion}.

\section{DESIGN} 
\label{sec:design}
\subsection{Science drivers} 
\label{subsec:scidri}
These IFUs are designed for studying nearby galactic stellar populations, the ISM of other galaxies, and stellar and gas kinematics of nearby galaxies.
Science drivers that influenced the design of the IFUs include:
probing the vertical structure of spiral disks using stellar and gas kinematic tracers;
studying the kinematics and abundances of diffuse ionized gas in edge-on and face-on spiral galaxies;
understanding the origin of winds in starburst galaxies;
measuring the rate of galactic outflows in normal spirals;
and connecting the properties of galaxy cores to the secular evolution of their parent galaxies for E+A galaxies, QSO/AGN hosts, and pseudo-bulges.

\subsection{Head design} 
\label{subsec:heads}
The primary limitations on the overall head size result from technical limitations in the telescope and spectrograph system.
The maximum number of fibers is limited simply by the maximum allowable slit length.
The overall field of view is limited by the IFU mount on the telescope.
The IFU heads mount into the WIYN Fiber-Optic Echelle (WIFOE) port on the telescope.
The WIFOE port limits the overall head mount to a 1in circular diameter.
Accounting for mounting hardware, this limits the overall physical IFU head size to approximately a square 0.5in$\times$0.5in, corresponding to a maximum field of view (FOV) of $119\arcsec\times119\arcsec$ at the WIYN plate scale of 9.\farcs374/mm.
Packing fibers and head mount fixturing make the usable FOV somewhat smaller than this, as shown later.

Each IFU head design presented a unique challenge for packing the fibers together to form the heads.
The main problem we needed to overcome was one of ``circle packing'': what is the best way to arrange the fibers (represented as circles as viewed along the optical axis of the system) so that we maximize the focal plane filling factor in a compact region while achieving our target scientific capability?
To our advantage, the problem of circle packing has been relatively well-studied in the field of geometry.
For a single diameter, the highest density packing arrangement is a hexagonal lattice arrangement as used in both DensePak and the SparsePak array on WIYN, which results in a packing density of $\pi/\sqrt{12} \approx 0.907$ \cite{Steinhaus99}.
Circle packing becomes markedly more difficult when including more than one circle diameter in the packing, as each of these new IFU heads does.
Our conceptual goal for these IFUs was to enable the simultaneous sampling of varying surface brightness levels in the same IFU observation.
To that end, the challenges of circle packing were central to our ability to design these IFU heads in a way that achieved our scientific goals yet were within our ability to fabricate.
In the following sections we describe the challenges we have overcome to develop scientifically-useful head designs that were also feasible to construct.

\subsubsection{HexPak}
\label{subsubsec:hexhead}

\begin{figure}
    \includegraphics[width=0.48\textwidth]{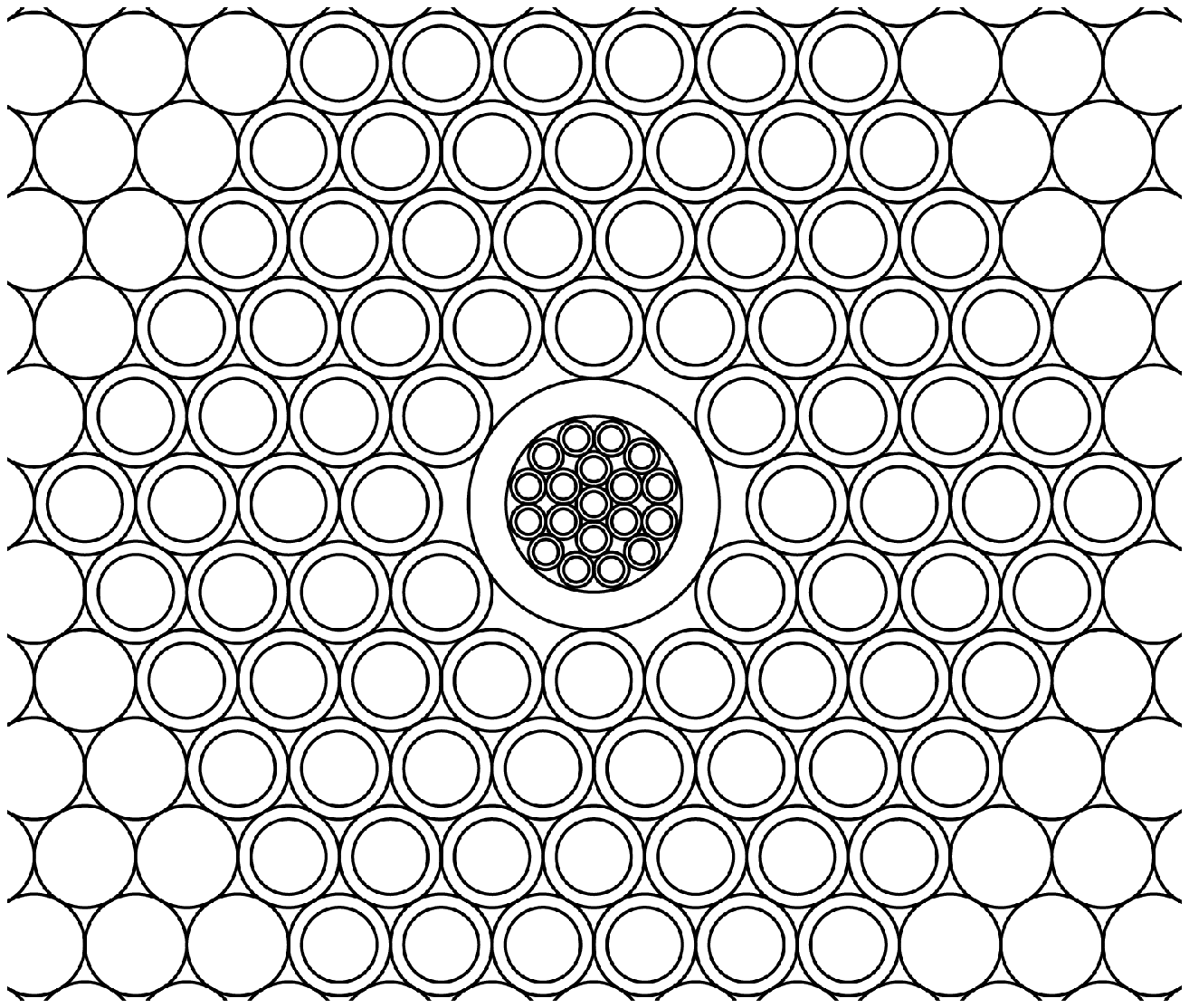}
    \hfill
    \includegraphics[width=0.48\textwidth]{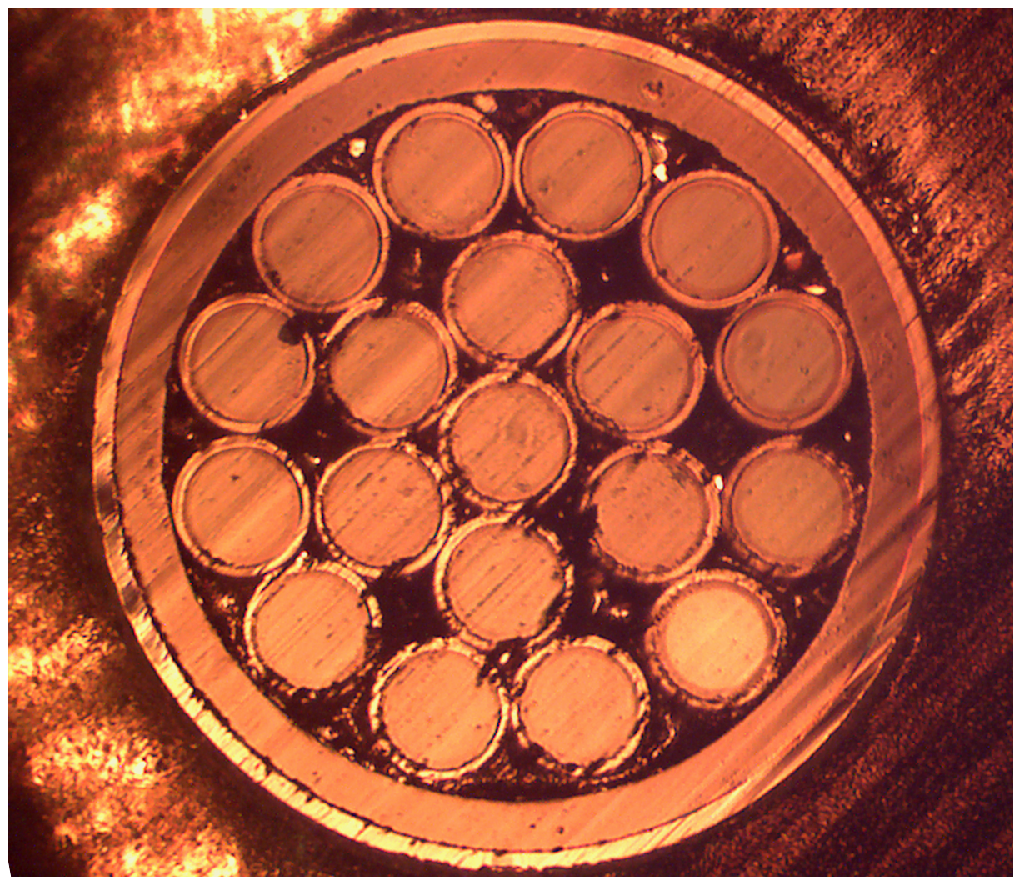}
    \vspace{1.5mm}
    \caption{
    \emph{Left:} Detail of the HexPak science fibers.
    Fiber core regions are not shown for packing fibers.
    The region depicted is approximately 0.18in$\times$0.15in centered on the hexagon.
    \emph{Right:} Prototype of the HexPak core packing.
    The fibers and capillary were glued into a 0.25in brass ferrule and ground to an even length in order to observe the quality of the packing.
    This face has not been well-polished and does not represent our final target level of surface polish.
    \label{fig:hex_head}}
\end{figure}

The HexPak IFU head is designed to study objects with radial surface brightness profiles (e.g.\ face-on galaxy disks, spheroidal galaxies).
As the name suggests, HexPak is a hexagonal fiber array, based on a hexagonal lattice arrangement.
As a result, the larger-diameter HexPak fibers are arranged in the most compact manner possible.
The fiber diameter for the hexagonal region was set by the diameter of the pre-existing DensePak fibers.
These fibers have core diameters of 310$\mu$m and full outer diameters measured to be $405\mu$m.
At the plate scale of the WIYN focal plane the cores of these fibers span 2.\farcs8\ on the sky.
The design goal for the HexPak head was to have a hexagonal region of fibers in an annulus around a high-spatial resolution fiber core, necessitating two different fiber diameters.
There are only nine possible compact\footnotemark\ packings of two circle diameters in a plane \cite{Kennedy06}.
\footnotetext{A packing is ``compact'' if, for every circle \emph{C} that is tangent to a series of circles \emph{C}$_1$,\emph{C}$_2$,\dots,\emph{C}$_n$, circle \emph{C}$_i$ is tangent to circle \emph{C}$_{i+1}$ for $i = 1,2,\dots,n$.}
Of these, only one is particularly relevant to the science goals of HexPak.
This packing involves replacing one larger circle with an array of seven smaller circles (with a diameter ratio of $\sim$0.386) while retaining the hexagonal lattice arrangement of the other large circles (see Figure~1.e in Ref.~\citenum{Kennedy06}).

The main flaw in this design as it pertains to our design goals is that the seven smaller circles \emph{must} be surrounded by larger circles to maintain a compact packing.
Stated differently, this arrangement dictates that the high-resolution core of HexPak be no larger than the diameter of one large fiber.
The reason behind this is that the largest radius occupied by the seven smaller fibers is larger than the radius of the one fiber which was removed.
Adjacent packings of seven small fibers would therefore overlap each other.
This means that this packing arrangement would allow for higher spatial sampling of no larger than a 2.\farcs8\ diameter region, but we felt a larger region would more closely meet our science goals.
In order to achieve a larger high-resolution core, we realized that this particular compact packing can be reversed by instead replacing seven circles with one larger circle.
This reversal ultimately led to the final design.

For the final HexPak head design we have replaced the central seven 2.\farcs8\ fibers in the hexagon with a glass capillary tube.
The outer diameter of this tube is such that it packs compactly to the surrounding fibers, and the inner diameter of the tube leaves a large, circular profile in which we can pack small fibers.
Our final task was to determine how to densely pack a moderate number of fibers inside this new circular aperture.
In order to determine an efficient packing within this aperture, we once again turned to the study of circle packing.
A sub-field of circle packing investigates the most efficient methods for packing circles of a given diameter into one circle of a larger diameter.
Ref.~\citenum{Kravitz67} presents empirically-derived optimal packings of circles into a circular container.
For a diameter ratio of $\sim$0.206, it is possible to pack 19 small circles into one larger circle.
This final design of all the HexPak object fibers can be seen on the left in \green{Figure~\ref{fig:hex_head}}.

It was prohibitively expensive to fabricate a form for a custom diameter capillary draw just a few inches in length, so in practice our chosen diameter (and the diameters of the fibers in the core) was limited to stock sizes.
In the case of HexPak, our chosen glass capillary was purchased from Polymicro Technologies\footnotemark\ and has a specified inner diameter of 750$\mu$m.
\footnotetext{Polymicro Technologies, 18019 N.\ 25th Avenue, Phoenix, AZ 85023--1200, (602) 375--4100}
This capillary inside diameter corresponds to a core fiber outer diameter of about 154$\mu$m.
The closest stock fiber outer diameter from Polymicro was 140$\mu$m (100$\mu$m or 0.\farcs94 core diameter), resulting in the final design of 19 0.\farcs94\ fibers packed in the center of the hexagon.
One sacrifice we make in using stock sizes is that there is a 2--3\arcsec\ gap between the core fibers and the larger fibers in the hexagonal region.
The necessary outer diameter to fill the space of seven 405$\mu$m OD fibers is 1,050$\mu$m, but this capillary has an outer diameter of 850$\mu$m. 
This amount of undersizing is problematic for the packing arrangement.
We manually thickened the pieces of capillary used in the head by applying thin coatings of spray paint to the tubes until we reached the desired diameter.

\begin{figure}[t]
    \centering
    \includegraphics[width=.55\textwidth]{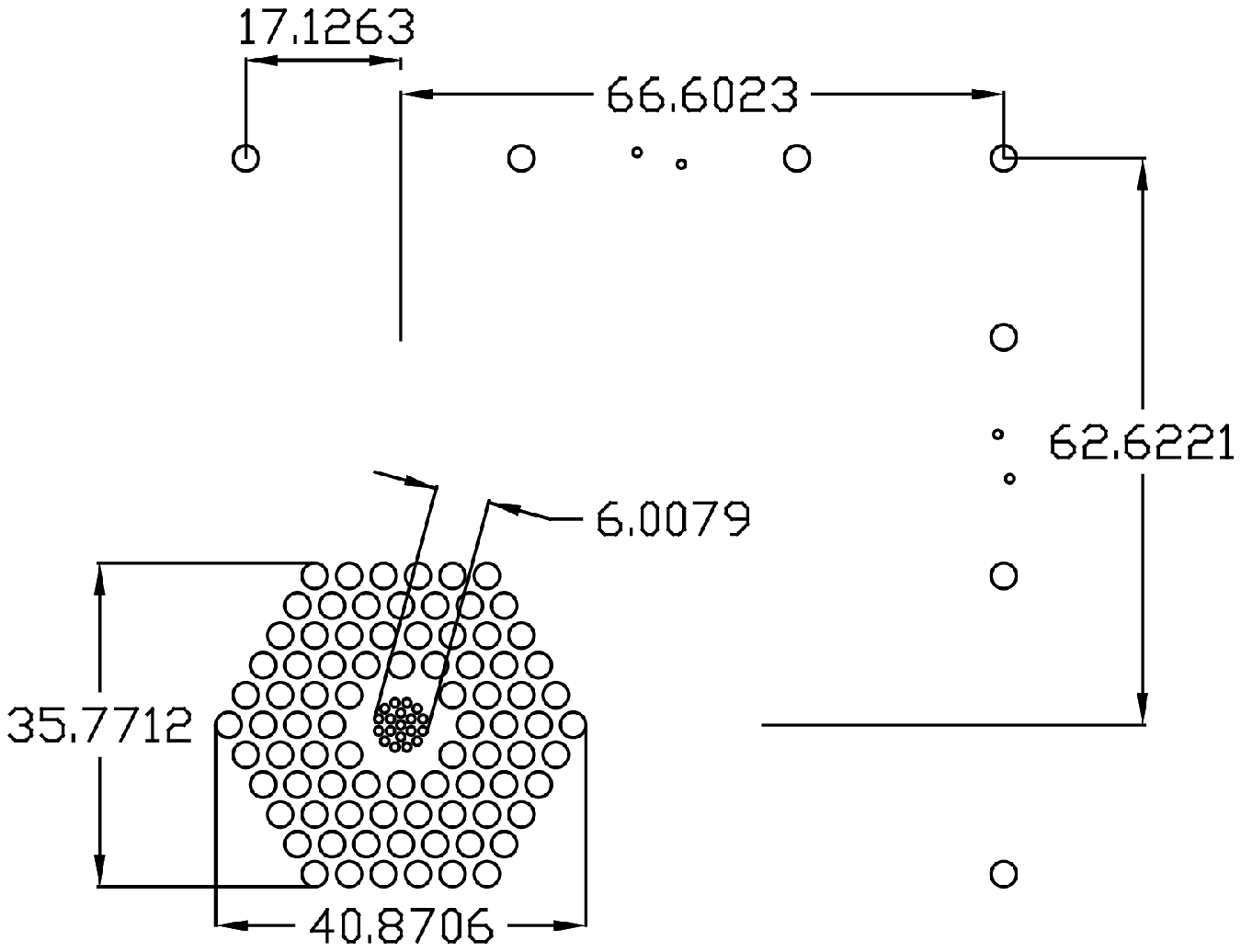}
    \hfill
    \includegraphics[width=.42\textwidth]{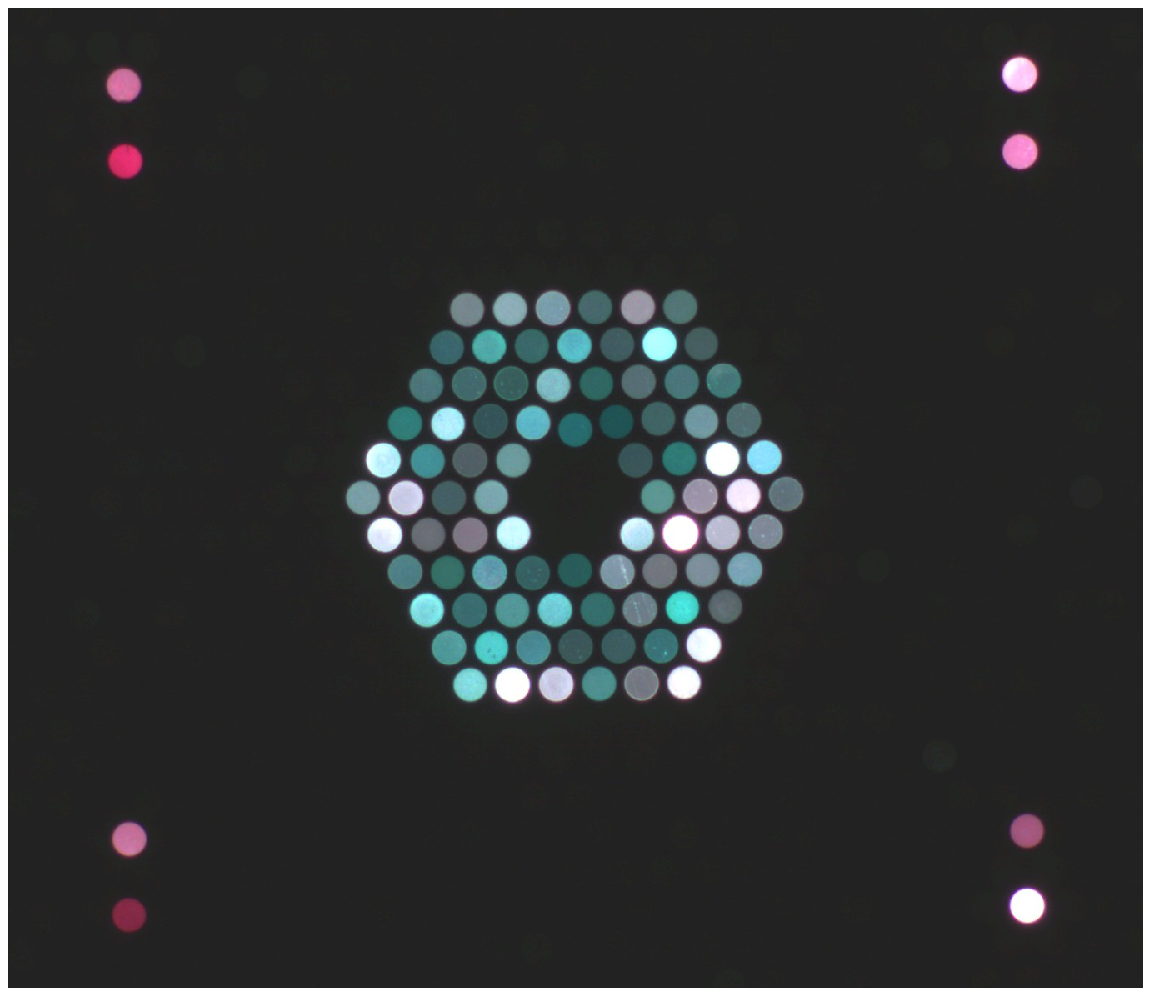}
    \vspace{1.5mm}
    \caption{
    \emph{Left:} Head design for the HexPak IFU.
    All displayed values are in units of arcseconds.
    The two different fiber diameters are 2.\farcs8 and 0.\farcs94 on the sky.
    Only the active fiber core regions are shown.
    \emph{Right:} An image of an early prototype of the HexPak head design.
    The center of the hexagonal region contains a hypodermic needle as a placeholder for the capillary with small fibers.
    No small fibers are in the hypodermic shown here.
    The ends of the fibers were colored with marker to easily differentiate object fibers (green) from sky fibers (red).
    \label{fig:hexpak}}
\end{figure}

Although the results from Ref.~\citenum{Kravitz67} are empirical (and therefore may not be compact) and our chosen fibers are slightly undersized, the amount of extra space in the packing arrangement is at an appropriate level to allow for mechanical tolerances. 
We have successfully assembled two prototypes of this packing and have high confidence that this will be viable in the final head assembly.
The right side of \green{Figure~\ref{fig:hex_head}} shows one of these prototypes after being glued and roughly polished.

Our final head design is shown in \green{Figure~\ref{fig:hexpak}}.
The array design has 114 total fibers.
Of these, the main science fiber region consists of 84 fibers with 2.\farcs8\ diameters contained within the outer hexagonal region and 19 fibers with 0.\farcs94\ diameters spanning the central 6\arcsec\ diameter of the hexagon.
The capillary tube housing the 0.\farcs94\ fibers creates an annular gap 2--3\arcsec in radius between the core fibers and the outer hexagonal region.
The array also has 11 total sky fibers, seven with 2.\farcs8\ diameters and four with 0.\farcs94\ diameters.
The sky fibers are placed along the edges of the array opposite the science fibers to maximize the separation between the object and sky sampling regions.
The sky fibers lie between 45\arcsec\ and 75\arcsec\ from the \emph{edge} of the hexagon.
The HexPak head design employs nearly 1,000 short (3in) packing fibers to fill the space in between the science fibers and the sky fibers, similar to the scheme used for SparsePak.
The packing fibers also create a roughly rectangular exterior head profile.
An early prototype of the HexPak design is also shown in \green{Figure~\ref{fig:hexpak}}.
This prototype shows an earlier revision in sky fiber arrangement and contains no small fibers in the core of the array.

Ref.~\citenum{Bershady04} found that focal ratio degradation\footnotemark\ (FRD) increases at the edge of the SparsePak array.
\footnotetext{Focal ratio degradation is a decrease in focal ratio of a beam as it passes through a fiber.}
They suggest several possible causes, including stress incurred when releasing the head from its mold and pressure introduced by the head-mount clamp.
Regardless of the cause, they recommend leaving at least a 1.2mm buffer region at the outside of the head in order to protect the active fibers from these potential stresses.
The ``packing fiber'' buffer in the HexPak head is four large fibers, or about 1.6mm, which we expect will alleviate FRD increases at the edge of the array.

\subsubsection{GradPak}
\label{subsubsec:gradhead}

\begin{figure}[t]
    \centering
    \begin{minipage}[c]{0.5\textwidth}
        \includegraphics[width=0.95\textwidth]{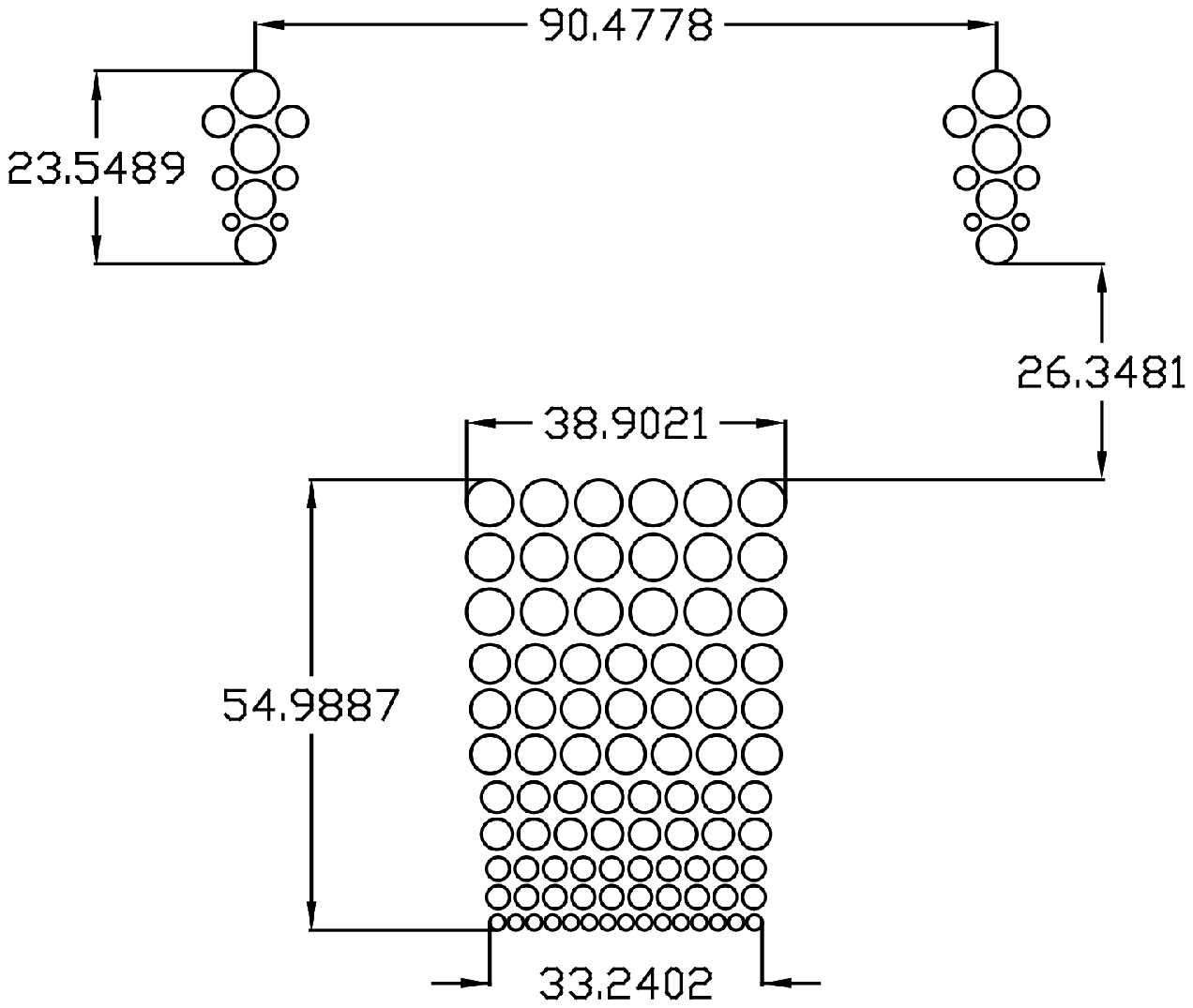}
    \end{minipage}
    \hfill
    \begin{minipage}[c]{0.41\textwidth}
        \includegraphics[width=0.95\textwidth]{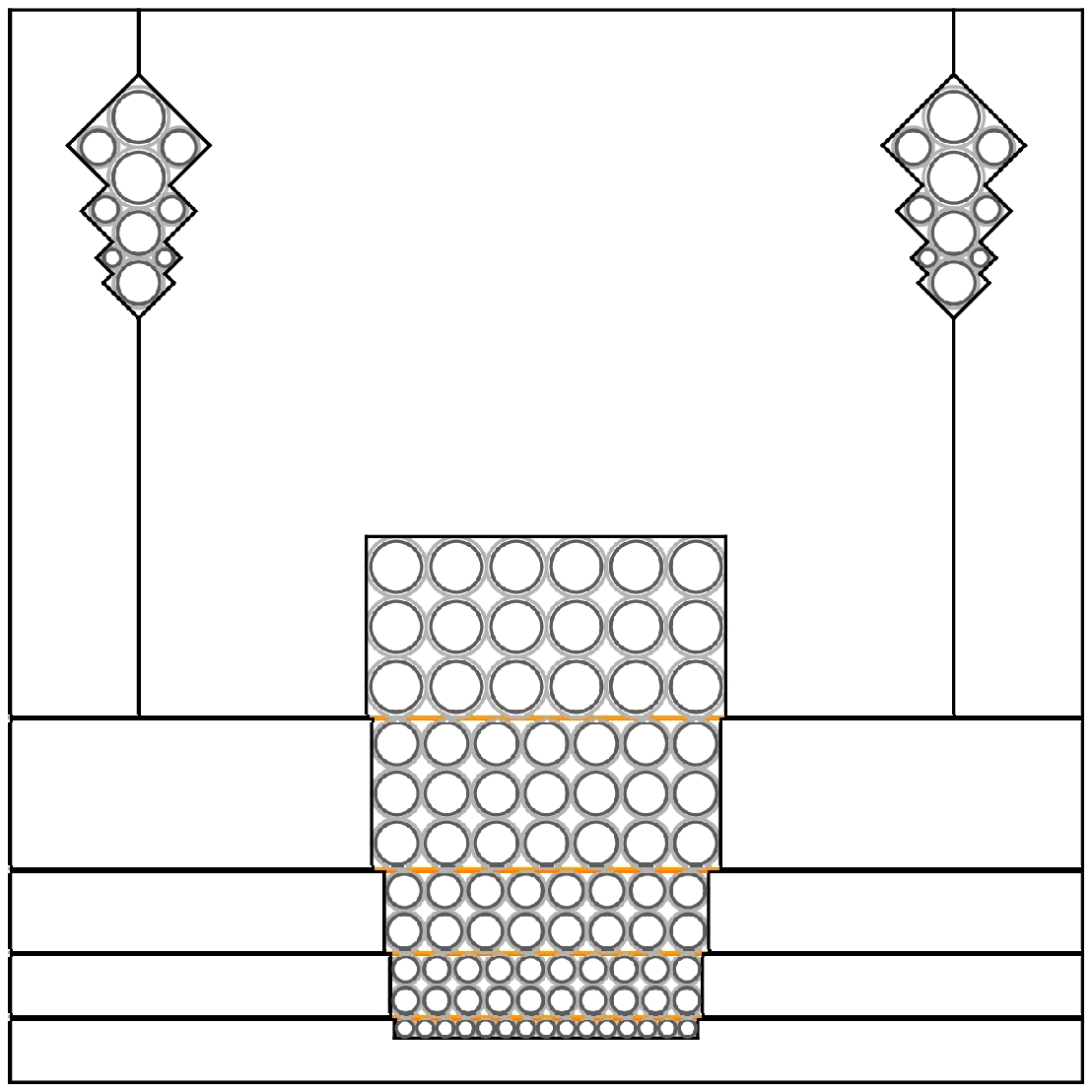}
    \end{minipage}
    \vspace{1.5mm}
    \caption{
    \emph{Left:} Fiber layout for the GradPak IFU.
    All displayed values are in units of arcseconds.
    The different fiber diameters are 1.\farcs9, 2.\farcs8, 3.\farcs7, 4.\farcs7, and 5.\farcs6 on the sky.
    Only the active fiber core regions are shown.
    \emph{Right:} Detail of the GradPak head mount.
    The darker gray circles are the active fiber cores, the lighter gray circles are the full O.D.\ of the fibers, and the orange lines represent $25\mu$m-thick plastic dividing layers.
    The full exterior profile measures 0.5in$\times$0.5in.
    \label{fig:gradpak}}
\end{figure}

GradPak is designed to study objects that exhibit linear gradients in surface brightness (e.g.\ edge-on galaxy disks, linear outflows).
Therefore, the desired fiber arrangement in the head is linear rather than approximately circular as it is in HexPak.
The design goal was to approximate layered pseudo-slits of increasing slit width, accomplished by using stacked rows of fiber, with fiber diameters increasing between rows.
There is no compact packing of two different circle diameters that allows for each fiber size to be arranged into linear, contiguous rows.
Circle packing theory did not show obvious solutions for this problem, so we settled on a design that uses five physically separated fiber regions, each containing only a single fiber diameter.
This design is shown on the left in \green{Figure~\ref{fig:gradpak}}.
Our final design includes a total of 110 fibers composed of five different fiber core diameters: 200$\mu$m, 300$\mu$m, 400$\mu$m, 500$\mu$m, and 600$\mu$m.
Respectively, these fibers span angular diameters of 1.\farcs9, 2.\farcs8, 3.\farcs7, 4.\farcs7, and 5.\farcs6\ at the WIYN focal plane.
In order to successfully accommodate five different fiber diameters within the same packing arrangement, we have designed a layered head mount.

This head mount fixturing represents one of the largest engineering challenges we have overcome.
In order to get each row of fibers to remain parallel to each other, fibers of different diameters could not touch.
This is a result of there being no compact packing of two different fiber diameters that allows for parallel rows.
This meant that it would be impossible to use short packing fibers in the GradPak head.
After numerous iterations, we ultimately designed a layered head mount that is assembled in stages, one fiber-diameter region at a time.
The final head mount design can be seen on the right in \green{Figure~\ref{fig:gradpak}}.
Each set of fibers of a given diameter is contained by aluminum walls, precisely machined to the height of each fiber region.
The head is then assembled by layering subsequent stacks of these segments on top of one another.
All of the parts of the fixture are held together as an assembly using small dowel pins that run the full height and width of the assembly.
The main cavity was cut out as an assembly using a wire electric discharge machining (EDM) process to ensure the best possible alignment between adjacent fiber regions.
The GradPak head block cavity was cut using a 0.004in diameter wire, allowing for a minimum corner radius of 0.002in (50$\mu$m), small enough to ensure the smallest fibers fit into place.

The dimensions of the entire head block are 0.5in$\times$0.5in.
Each region of fibers containing a single fiber diameter is separated from its neighbors by a thin layer of plastic shim stock 0.001in (25.4$\mu$m) thick.
Although this adds a gap between each region of fibers, the on-sky angular size of the plastic layer is very small (0.\farcs24), which we deemed acceptable.
We have successfully designed and assembled two prototype GradPak head assemblies and we are confident that this design is feasible.

The resulting array design consists of five stacked blocks of fibers arranged into one or more rows, with the fiber core diameter increasing from 200$\mu$m to 600$\mu$m between each block of fibers.
The entire block of science fibers includes 90 fibers and covers an area on the sky roughly 35\arcsec$\times$55\arcsec.
The array also has two regions of sky fibers for simultaneous sky measurements.
Each sky fiber region contains two fibers of each fiber diameter included in the science fiber region.
The fibers in GradPak are arranged in a square lattice rather than in a hexagonal lattice, reducing the effective filling factor within each fiber region.
This yields a packing density of only $\pi/4 \approx 0.785$.
The advantage of this arrangement is that the pseudo-slits within each region are vertically aligned and are offset only in one dimension on the sky.

Since the entire GradPak head mount assembly is composed of aluminum parts, no short packing fibers are used in its design.
The design does, however, place active fibers within 1.2mm of the edge of the head, inside the prescribed minimum buffer distance.
However, we believe FRD edge effects should be negligible due to the use of a stronger buffer material (aluminum in this case) and the fact that at no point during assembly is any part of the GradPak head released from a mold.


\subsection{Slit design} 
\label{subsec:slit}

\begin{figure}[t]
    \centering
    \includegraphics[width=\textwidth]{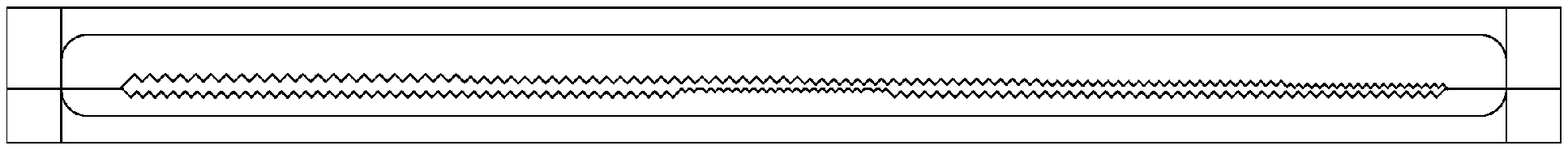}\\
    \vspace{2pt}
    \includegraphics[width=\textwidth]{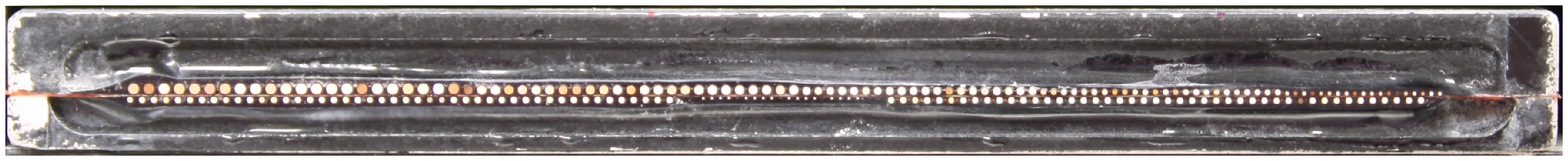}
    \caption{
    \emph{Top:} A schematic view of the slit block without fibers.
    Both halves of the slit block assembly are shown.
    The bottom set of grooves is the HexPak slit and the upper set of grooves is the GradPak slit.
    \emph{Bottom:}
    An image of the slit face in its current state.
    The top row of fibers are GradPak, arranged by decreasing diameters.
    The bottom row of fibers are HexPak, with the 0.\farcs94 fibers situated between the 2.\farcs8 fibers.
    Brightness differences between fibers are largely due to surface irregularities in the unpolished head end.
    Color differences are simply due to some fibers pointing at different parts of the lab.
    The two 0.001in-thick plastic shim stock dividers are seen as a thin orange line separating the two slits.
    The shown view spans 3.562in$\times$0.312in.
    \label{fig:slit}}
\end{figure}

GradPak and HexPak both terminate in the same ``foot'' housing in a dual-slit configuration.
The slit block uses a custom design that modified the SparsePak slit block design.
The slit block assembly is made from two separate aluminum plates, each specifically designed to create a pseudo-slit for one of the IFUs.
Each pseudo-slit was created by cutting `v'-grooves into the slit block.
The groove size, placement, and spacing is designed to match the varying number and sizes of fibers in each IFU pseudo-slit.
A schematic of the slit block assembly is shown in \green{Figure~\ref{fig:slit}}.
For mechanical reasons, the fibers are aligned to be tangent to the optical centerline of the slit, rather than being aligned by their centers.
The `v'-grooves were precision-cut into the slit block using a wire EDM process like that used for the GradPak head mount fiber cavity, but with a smaller $0.002$in diameter wire, resulting in a minimum corner radius of approximately $0.001$in ($25\mu$m).
This radius is smaller than the smallest fiber outer radius (measured to be about 71$\mu$m) ensuring that even the smallest fibers will sit nicely into the grooves.

We have set the fiber core edge-to-edge spacing at $291\mu$m to minimize cross-talk while maximizing the number of fibers in the slit.
Ref.~\citenum{Bershady04} found that 400$\mu$m was the ideal edge-to-edge spacing of fiber cores in the slit in terms of the trade-off between cross-talk and packing density (i.e.\ total number of fibers).
However, this determination was made using the old Bench Spectrograph system which was upgraded in 2008\cite{Bershady08}.
With improved image quality and detector sampling as a result of the upgrade, we estimate we can reduce the fiber spacing by $\sim25\%$ without introducing significant amounts of cross-talk between fiber channels.

The fiber arrangements in the slit are informed by their locations within the head design.
In HexPak, the hexagonal region is divided into six equal wedges, each containing 14 fibers.
The fibers within each wedge are located in the same region of the slit, with fibers near the core of the array being located in the center of the slit segment.
Each wedge section in the slit is separated from the adjacent wedge sections by a sky fiber, allowing sky sampling across the entire slit length.
Three wedges and associated sky fibers are located on each end of the slit, separated by all of the 0.\farcs94 fibers.
The 0.\farcs94 fibers are located in the center of the slit, with the associated 0.\farcs94 sky fibers being located on either side of the 0.\farcs94 object fibers.
The GradPak slit is sorted by fiber diameter.
Within each diameter region, each fiber row in the head is placed contiguously in the slit, with sky fibers evenly distributed as separators between rows.

We also slightly modified the standard Bench Upgrade design for the ``toes'' of the cable in order to negate vignetting of the light exiting the fiber slit.
The toes house the slit block and have open chambers for inserting blocking filters, slit masks, and a back-illumination mirror.
These chambers each have an aperture through which the exit beam must pass in order to reach the spectrograph collimator.
We performed a vignetting analysis to determine the maximum fiber slit width and slit length that could pass unvignetted through the toes.
Ref.~\citenum{Bershady04} shows that, due to FRD, 99\% of the energy in an \f6.3 input beam (the WIYN focal ratio) exiting the fibers is contained within an \f4 output beam.
Therefore, as long as an \f4 beam could pass through the toes without being blocked, we could consider light losses from the toes to be negligible.
The SparsePak design for the toes allowed for a maximum slit length of 79.8mm under this vignetting requirement.
This number informed the fiber-to-fiber spacing for our final slit design, resulting in an overall slit length of 76.6mm.
The maximum allowable slit width for an \f4 beam was 1.23mm, or a $615\mu$m beam displacement from slit center.
Our largest fiber has a $600\mu$m core diameter with approximately $55\mu$m of clading and buffer in radius.
Each half of the slit block also has a $25\mu$m plastic divider at slit center, a requirement of our slit block construction process.
In total, the maximum beam displacement from slit center is about $680\mu$m, too large for the standard design for the toes.
As a result, we modified the design and increased the width of one critical aperture in the toes such that the maximum allowable displacement from slit center is $927\mu$m.

\section{CONSTRUCTION AND STATUS} 
\label{sec:construction}

The design and construction of the SparsePak cable served as the starting point for this instrument.
Technical details of SparsePak fabrication are available in the Appendix section of Ref.~\citenum{Bershady04}.
In this section we focus primarily on detailing the important features unique to this instrument and noting significant departures from the SparsePak design.

At the present time, construction is complete on all cabling and the entire spectrograph feed assembly (the foot).
Polishing of the fiber slit is nearly complete, and final gluing and polishing of the IFU heads are scheduled to conclude in June--August 2012.

\subsection{Fiber optics}
\label{subsec:fibers}
The 91 2.\farcs8\ fibers that compose the hexagonal region and its associated sky fibers are re-purposed from the DensePak IFU.
They have a specified active core diameter of $310\mu$m and a measured total O.D.\ (including cladding and buffer) of approximately $405\mu$m.
Ref.~\citenum{Barden98} states that the DensePak fibers are similar to the Hydra FIP-type ``red-optimized'' fibers.
The 23 0.\farcs94\ fibers forming the high-resolution core and associated sky fibers were purchased new from Polymicro Technologies.
These new fibers are FBP-type broad-spectrum optical fiber with excellent throughput for optical wavelengths.
The fibers have silica cores with silica cladding and are designed to have excellent throughput across the visible spectral window and into the near-infrared.
The diameter specification (given as core:clad:buffer, in microns) is 100:120:140.
This new fiber is very similar to the fiber used in SparsePak and therefore we expect similar throughput performance from the core HexPak fibers.
Throughput testing for SparsePak showed a roughly 20\% improvement in throughput compared to the DensePak fibers, after accounting for differences in fiber diameter and collected solid angle\cite{Bershady05}.
However, most of the light loss in the DensePak cable was likely due to vignetting in the original design of the fiber cable toes, before the Bench Upgrade.
With our modified version of the standard Bench Upgrade toes, we expect these fibers to have comparable performance to the SparsePak fibers in the red ($>4000$\AA; the FBP-type fibers have better transmission properties below 4000\AA\ compared to the FIP-type fibers).

The GradPak IFU consists entirely of new FBP-type broad-spectrum optical fiber from Polymicro Technologies.
The fibers have specified diameters (given as core:clad:buffer, in microns) of 200:220:240, 300:330:370, 400:440:480, 500:550:590, and 600:660:710.
This fiber is very similar to the fiber used in SparsePak and therefore we expect similar throughput performance from all the GradPak fibers.

\subsection{Cabling}
\label{subsec:cabling}
Each fiber is individually housed in a PTFE tube for protection and strain relief.
The PTFE tubes for the $310\mu$m HexPak fibers were also re-used from the DensePak cable.
All of the newly-purchased fiber is housed in new PTFE tubing from Zeus, Inc\footnotemark.
\footnotetext{Zeus, Inc., 3737 Industrial Blvd., Orangeburg, SC 29118, (800) 526--3842}
As a design decision to reduce the radial cable size, the new PTFE tubes have ``light-weight wall'' thicknesses (0.006in in radius).
This has proven troublesome due to the relative lack of radial strength and rigidity in the wall compared to thicker-walled tubing.
The very thin walls of these tubes are prone to kinking which threatens the integrity of the fibers they serve to protect.
We recommend, at minimum, ``thin wall'' PTFE tubes ($\gtrsim$0.01in).

The PTFE tubes terminate at each end of the cable in metal shear clamps.
The foot-end shear clamp contains an array of approximately $9\times27$ holes in a three-element aluminum clamp.
The elongated design of this clamp serves to transform the circular bundle of fibers into a rectangular shape approximating the slit arrangement.
The head-end shear clamps are a similar three-element design and function in a similar fashion to the foot shear clamp.
The hole pattern in the head end shear clamps is roughly hexagonal in order to accommodate the circular fiber bundle and to approximate the fiber layout of the HexPak head.

The fibers and their PTFE tubes are encased in 75 feet of flexible metal conduit.
The conduit housing the shared length of the cable consists of approximately 57 feet of 2in ID IE30 and IE50\footnotemark\ interlocking stainless steel exhaust hose.
\footnotetext{Penflex, Corp., 105B Industrial Drive, Gilbertsville, PA 19525, (800) 232--3539}
This conduit retains flexibility for handling while providing strength and durability for the length of cable that will be enclosed within the telescope structure.

The remaining 18 feet of the cable is housed in two separate lengths of 1.25in ID flexible aluminum standard electrical conduit.
Each length contains the remaining length of fiber for one of the IFU heads.
The lightweight construction and smaller diameter of this conduit allows for ease of handling for mounting the IFU heads into the telescope's Nasmyth port.
These two lengths of conduit are completely covered in polyolefin and PVC heat-shrink tubing for additional rigidity and abrasion resistance.
The smaller lengths of conduit are joined to the larger, stainless steel conduit through a custom-built merge collector from the automotive industry\footnotemark.
\footnotetext{Specialty Design Products, Inc., 11252 Sunco Drive, Rancho Cordova, CA 95742, (888) 778--3312}

The conduit for each IFU head contains two custom-designed low-profile rotation couplers.
Each coupler has $180^{\circ}$ of rotation about the optical axis, allowing each IFU head to rotate a full $360^{\circ}$ with the telescope instrument port rotator.
The two rotation couplers are spaced approximately $12$in apart along the cable to ensure that the full rotation is distributed along a sufficient length to avoid fiber stress.

\subsection{Fiber slit}
\label{subsec:slitconstruct}
For gluing the fiber ends into the slit grooves, we fabricated custom metal fixturing to bend the fibers 90$^{\circ}$ in the foot and hold them with clearance to mount the slit blocks.
We used EPO-TEK\footnotemark\ 354 heat-curing optical epoxy to bond the fiber ends to the slit block.
\footnotetext{Epoxy Technology, Inc., 14 Fortune Drive, Billerica, MA 01821, (978) 667--3805}
The gluing fixture held each fiber in its respective slit position and allowed us to seat all the fibers simultaneously.
We used a thin sheet of plastic to cover the exposed fiber and epoxy and held the fibers in place using an aluminum pressing block while the epoxy cured.
We glued each slit block separately and then constructed the final assembly from the two completed halves.
The two halves are pinned together to ensure accurate, repeatable positioning between the two slit halves.

\subsection{Fiber heads}
\label{subsec:headconstruct}
As tested with SparsePak and our HexPak and GradPak test head assemblies, we will use the same EPO-TEK 354 heat curing epoxy for assembling the IFU heads.
The HexPak head is assembled one row of fibers at a time, using short packing fibers to fill in the spaces around the active fibers.
The 0.\farcs94\ fibers are pre-assembled into their capillary tubes and then placed into the final assembly during the gluing process.
For constructing the HexPak head, we follow a design and process used for SparsePak, based on a concept from S.\ Barden (\emph{private communication}).
We will assemble the fibers in a micrometer-driven vice with the vice channel width set precisely to the width of the head.
We will then use a precisely-machined tamping tool, just slightly narrower than the width of the head, to pack the fibers into the channel mold.
The mold and the tamping tool are sprayed with a PTFE mold release\footnotemark[1]\ prior to gluing.
\footnotetext[1]{Miller--Stephenson Chemical Company, Inc., 55 Backus Ave., Danbury, CT 06810, (203) 743--4447}

The GradPak head is assembled one fiber region at a time, with all fibers of a single diameter being assembled at the same time.
The assembly is cured after each additional region is assembled.
In contrast to the HexPak construction, no vice is used for GradPak.
The head mount parts are machined to the precise cavity width to hold the fibers for each region, requiring only a pressing plate to ensure the fibers remain seated in the cavity while curing.
The pressing plate is coated in PTFE mold release as a precaution and separated entirely from the epoxy by the plastic dividing layers, so at no point are the fibers released from a cured mold.
This should alleviate any additional FRD introduced through fiber stresses when releasing the mold, as seen in SparsePak.

\subsection{Polishing}
\label{subsec:polishing}
We will polish the optical surfaces of the fiber slit and each of the fiber heads in order to minimize light losses and FRD as the light passes through the cable.
We are using an UltraPol\footnotemark[2]\ 1200 model lapping machine with 8in diameter silicon carbide and aluminum oxide lapping disks.
\footnotetext[2]{ULTRA TEC Manufacturing, Inc., 1025 E.\ Chestnut Avenue, Santa Ana, CA 92701, (877) 542--0609}
The lapping disks span a range of grit sizes, starting from $70\mu$m for the grinding phase to $0.3\mu$m for the final polishing phase.
Our laboratory testing shows that FRD is minimized for surface roughnesses $<5\mu$m.
Our target surface roughness is $<0.5\mu$m.
See Eigenbrot et al.\ in these proceedings for a full description of our FRD and surface polish testing.
At the time of this writing, the entirety of the slit was polished at a 5$\mu$m level and ready to proceed to 1$\mu$m grit disks, followed by 0.3$\mu$m grit.
An image of the slit in this state is shown in \green{Figure~\ref{fig:slit}}.

\section{SUMMARY} 
\label{sec:conclusion}

We are in the final construction phase of two new fiber optic IFUs, GradPak and HexPak.
These IFUs will be the first formatted fiber IFUs to utilize multiple fiber diameters within the same IFU head.
By including multiple fiber diameters these IFUs will greatly expand the spectroscopic capabilities of the WIYN 3.5m telescope, providing the ability to sample varying spatial scales within the same observation with the highly versatile Bench Spectrograph.
This will enable observations simultaneously spanning a wide range in surface brightness to be optimized for the photon limit at spectral resolutions between 1000 $< \lambda/\Delta\lambda <$ 30,000.

HexPak is designed to observe axi-symmetric surface brightness profiles with a $36\arcsec\times41\arcsec$ hexagonal region sampled by 2.\farcs8\ fibers and a 6\arcsec diameter high-resolution core sampled by 0.\farcs94\ fibers.
GradPak is optimized for linear surface brightness gradients, with a stacked pseudo-slit design spanning $39\arcsec\times55\arcsec$ using fibers ranging from 1.\farcs9\ to 5.\farcs6\ in diameter.
Each of these IFUs presented unique challenges for successfully incorporating multiple fiber diameters within the same fiber head.
We have described two methods for overcoming these challenges, one for radial arrangements of fibers and one for linear arrangements.
Our solutions optimize the packing fraction of science fibers while also achieving regular and precise placement and configuration of fibers within each IFU.



\section*{ACKNOWLEDGMENTS}
The authors would like to thank Charles Corson, Pat Knezek, and George Jacoby for their support of this project at WIYN.
This research was supported by NSF/ATI-0804576 and NSF/AST-1009471. 


\bibliography{8446-106}   
\bibliographystyle{spiebib}   

\end{document}